\documentclass[twocolumn,superscriptaddress,floatfix,prl,showpacs]{revtex4}
\usepackage[dvips]{graphicx}

\begin{document}

\title{\Huge Polaronic metal in lightly doped high-T$_c$ cuprates \\
\mbox{} \\}

\author{\Large A.~S.~Mishchenko}
\affiliation{Cross-Correlated Materials Research Group, RIKEN,
2-1 Hirosawa, Wako, Saitama, 351-0198, Japan}
\affiliation{RRC ``Kurchatov Institute", 123182, Moscow, Russia}

\author{\Large N.~Nagaosa}
\affiliation{Cross-Correlated Materials Research Group, RIKEN,
2-1 Hirosawa, Wako, Saitama, 351-0198, Japan}
\affiliation{Department of Applied Physics, The University of Tokyo,
7-3-1 Hongo, Bunkyo-ku, Tokyo 113, Japan} 

\author{\Large K.~M.~Shen}
\affiliation{Laboratory of Atomic and Solid State Physics, Department of Physics, Cornell University, Ithaca NY 14853, USA}

\author{\Large Z.-X.~Shen}
\affiliation{Department of Physics, Applied Physics, and
Stanford Institute for Materials and Energy Sciences, 
SLAC National Accelerator Laboratory, Menlo Park, CA 94025 and
Stanford University, Stanford, CA 94305 USA}

\author{\Large X.~J.~Zhou}
\affiliation{National Lab for Superconductivity, Institute of Physics, Chinese Academy of Sciences, Beijing 100190, China}

\author{\Large T.~P.~Devereaux}
\affiliation{Department of Physics, Applied Physics, and
Stanford Institute for Materials and Energy Sciences, 
SLAC National Accelerator Laboratory, Menlo Park, CA 94025 and
Stanford University, Stanford, CA 94305 USA}


\mbox{} \bigskip

\begin{abstract}
{\mbox{} \newline \mbox{} \newline \mbox{} \newline 
\Large We present a combined study of the 
angle-resolved-photoemission 
spectroscopy (ARPES) and quantum Monte Carlo 
simulations to propose a novel polaronic metallic state
in underdoped cuprates.
An approximation scheme is proposed to represent underdoped
cuprates away from 1/2 filling,
replacing the many-body Hamiltonian by that of a single polaron with
effective electron-phonon interaction (EPI), that successfully
explains many puzzles such as a large momentum-dependent
dichotomy between nodal and anti-nodal directions, and an unconventional
doping dependence of ARPES in the underdoped region. }
\end{abstract}

\maketitle


It is established that the physics of high temperature
superconductors is that of hole doping a Mott insulator \cite{Dag94,Lee}
where even a single hole is substantially influenced
by many-body effects \cite{Kane89}. 
Understanding of the dynamics of holes in Mott insulators
has attracted a great deal of 
interest \cite{Dag94,Kane89,Man91}. The major interactions
are electron-electron interactions (EEI) and EPI. The
importance of the former is no doubt essential since
the Mott insulator is driven by this EEI, while the latter was
considered to be largely irrelevant to superconductivity based on
the observations of a 
small isotope effect on the optimal $T_c$ \cite{isotope} and an
absence of a phonon contribution to the resistivity
\cite{resistivity}. 

On the other hand,  there is now accumulating evidence
that the EPI plays an important role in the physics of cuprates 
\cite{Gun08,UFN09}.
In particular, EPI manifests itself in
(i) an isotope effect on superfluid density $\rho_s$ and T$_{c}$
away from optimal doping \cite{Keller}, and (ii)  neutron
and Raman scattering \cite{Pint99,Raman1,Raman2}
experiments showing strong phonon softening with both
temperature and hole doping, indicating that EPI is strong
\cite{Khal97}.
Furthermore, the recent advancement in the energy/momentum resolution 
\cite{Shen_03} of ARPES 
resulted in the discovery of the dispersion "kinks" at around 40-70meV 
measured from the Fermi energy, in the range of the relevant 
oxygen related phonons \cite{Lanzara01,DevCuk2004a,DevCuk2004b}. 
These particular phonons - oxygen buckling and half-breathing modes are 
known to soften with doping \cite{Egami,Pint99} and with temperature
\cite{DevCuk2004a,DevCuk2004b,Egami,Pint99,Raman1,Raman2} 
and there is mounting evidence relating the kink to the phonon anomaly
of the bond stretching phonons \cite{Graf08,EPL10}. 
The quick change of the velocity can be predicted by any interaction of
a quasiparticle with a bosonic mode, either with a phonon because of EPI
\cite{DevCuk2004a,DevCuk2004b,Sangio10,EPL10} or 
with a collective magnetic resonance mode
\cite{ChuNor,EschNor}. 
Early studies of the doping dependence of the kink revealed a ``universality''
of the kink energy for La$_{2-x}$Sr$_x$CuO$_4$ (LSCO) 
over the entire doping range
\cite{Zhou03} and casted doubts on the validity of the latter scenario as the
energy scale of the magnetic excitation changes strongly with
doping. 
More recent studies showed that there is a subtle doping dependence of the 
kink energy \cite{Kordyuk06} which, nevertheless, can be explained
within the framework of EPI scenario as well \cite{Sangio10}.  

As a more direct test distinguishing between these two scenarios, an 
observation of an isotope effect just in the vicinity
of the kink \cite{DessauCM} and its explanation in terms of EPI
\cite{Koikegami08} has given an additional argument in favor of a phononic
origin of the kink.  
Furthermore, it was shown that the kink is observed in electron-doped cuprates
around 40-70meV, i.e. at the same energies as the hole-doped ones 
\cite{KimLambda}.
Hence, it is clear that the kink is due to EPI with phonon modes, which 
are roughly at the same energies in hole- and electron-doped cuprates, and not 
because of coupling to a magnetic mode whose energy in the electron-doped 
compounds is not larger than 10 meV \cite{Wilson06,Zhao07}. 

Generally, there are two possible pictures of the EPI. 
One is the {\it Migdal-Eliashberg picture} where the
kinetic (Fermi) energy of the electrons $\varepsilon_F$ is much
larger than the phonon energy $\Omega$. In this limit, the
multi-phonon processes represented by the vertex correction are
suppressed by the Fermi degeneracy and reduced by the adiabatic
factor $\Omega / \varepsilon_{F}$, i.e., EPI is basically in the
weak coupling region\cite{Migdal,Mahan}. The other limit is the
{\it polaron picture} where $\varepsilon_{F}$ is much smaller than
$\Omega$ where the multi-phonon processes can, in principle, lead 
to the small polaron formation.
The latter picture of the strong coupling limit of a polaron in {\it undoped} 
materials, where ARPES corresponds to the single hole dynamics in 
a Mott insulator, has been established  by a detailed comparison 
between experiment \cite{Kyle_Dop,ZX95,KyleTdep07} and theory 
\cite{tJph,RoGu2005,Rosch,CataTdep07}.
The picture obtained there is that the ``quasi-particle'' peak
observed experimentally is that of Franck-Condon multiphonon band,
which follows the energy dispersion of bare t-J model 
without EPI, while the zero-phonon line has only a very small
weight $Z<<1$. Namely the hole polaron is in the strong
coupling small polaronic state. 

At very {\it small dopings} it is clear that the Fermi energy 
$\varepsilon_{F}$ of a few holes doped into the Mott insulator
is smaller than the relevant phonon energy $\Omega$. 
Then the polaronic energy dispersion can be inverted
and the ARPES dispersion is cut-off by the (small)
Fermi energy $\varepsilon_{F}$ (measured from the hole side). 
In this case the adiabatic factor $\Omega / \varepsilon_{F}$ is not small and
one cannot rely on the weak coupling Migdal-Eliashberg approach and, hence, 
has to treat the EPI in the polaron limit including all vertex corrections and 
without considerable approximations.      
The polaron paradigm is complementary to the conventional 
Migdal-Eliashberg picture for metals \cite{Migdal,Mahan},
the latter of which fails to explain various anomalous features in 
the underdoped region.
Therefore it is reasonable to apply the polaron picture 
in the underdoped region.

In the present paper we interpret the kink in underdoped cuprates as 
a result of a short-range Holstein-like interaction of a hole
with optical phonons. 
Since the polaron scenario is more appropriate in the underdoped region
than the Migdal-Eliashberg one,   
we approximate the interacting many-body Hamiltonian by the single hole
polaronic Hamiltonian with an effective dimensionless coupling constant 
$\lambda$ of EPI.
The effective coupling constant $\lambda$ in the weakly doped compound 
is renormalized in comparison with the bare constant $\lambda_0$, 
corresponding to the case of a single hole, by EEI. 
It is known for gas of Fr\"{o}hlich polarons 
both theoretically and experimentally 
\cite{Tempere1,Tempere2,Mech,Dev,Meev} that the basic
properties of polarons behave with increasing concentration as if 
the coupling constant of the EPI is effectively reduced.
Comparison with experiment shows that effects of weak EPI and Coulomb 
repulsion can be factorized in 2D systems in the light doping limit and,
besides, this factorization is approximately valid up to a  
doping level $\delta \approx 0.1$ \cite{Tempere1,Tempere2}. 
Since the radius of Holstein polarons is smaller than that of Fr\"{o}hlich
ones we expect in our case better applicability of the modeling of the 
influence of doping by changing the effective EPI constant $\lambda$.  

The above assumption indicates that 
 $\lambda$ should be regarded as the effective coupling 
constant in the trial Hamiltonian which best mimics the experimental results.
Such an approach has been  already applied to the explanation of the doping 
dependence of the optical absorption spectra of underdoped cuprates 
and it was shown that the effective coupling constant $\lambda$ 
decreases with increasing doping \cite{tJOpt08}. 

\begin{figure*}
\includegraphics[width=19cm]{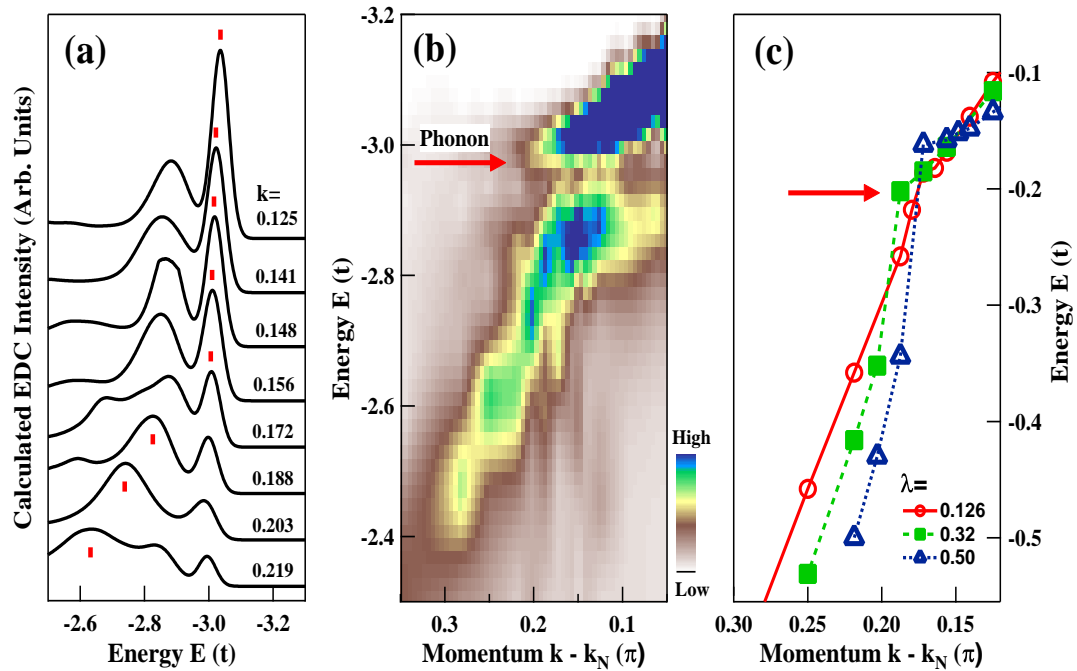}
\caption{\large (a) Evolution of the Lehman function along
the nodal $(\pi /2 , \pi / 2) \to (0,0)$ direction for $\lambda=0.5$.
Thick red lines indicate maxima of the Lehman function obtained 
from EDC analysis.
(b) Intensity plot for the same parameters. 
Energy in panels (a-b) is counted from the vacuum state of the
Hamiltonian (\ref{tJ}-\ref{e-ph}).  
(c) Momentum dependence of the peaks in the Lehman function for
$\lambda=0.126$ (circles), $\lambda=0.32$ (squares), and
$\lambda=0.50$ (triangles). 
Energy in panel (c) is counted from the top of polaron band at 
${\bf k}_N = (\pi /2 , \pi / 2)$.
The red arrow in (b) and (c) indicates the position of the phonon 
measured from the top of the polaron band.
All energies are in the units of $t=0.4$eV.}
\label{fig:fig1}
\end{figure*}

Using the approximation-free Diagrammatic Monte Carlo (DMC) method 
\cite{tJph,MPSSa,MPSSb,MPSS1a,MPSS1b} 
we have calculated the momentum dependence of
the Lehman spectral functions of a single hole in the extended 
$t-J$ model, i.e., $t-t'-t''-J-ph$ model (Fig.~\ref{fig:fig1}). 
Within this model a single hole in an
antiferromagnet can hop up to the 3rd nearest neighbor ($t(1)
\equiv t$, $t(2) \equiv t'$, and $t(3) \equiv t''$):
$$
\hat{H}_{\mbox{\scriptsize tt't''-J}} = -
            \sum_{n=1}^{3} \sum_{\langle ij,n \rangle \sigma}
            t(n) c_{i\sigma }^{\dagger} c_{j\sigma }
$$
\begin{equation} 
                + J \sum_{\langle ij \rangle}
\left( {\bf S}_i {\bf S}_j - n_i n_j / 4 \right) . 
\label{tJ}
\end{equation}
Here $c_{j\sigma}$ is a projected (to avoid double occupancy) fermion
annihilation operator, $n_i
=\sum_{\sigma}c^{\dagger}_{i\sigma}c_{i\sigma} < 2$ is the
occupation number, ${\bf S}_i$ is spin-1/2 operator, $\langle ij, n
\rangle$ denotes neighboring sites on $n$-th coordination sphere
(circle) in a two-dimensional lattice. 
We note, that although long-range antiferromagnetic (AFM) 
order and AFM gap is destroyed at 
rather low doping levels, overdamped spin waves and short range 
AFM correlations, showing the largest interplay with 
short-range EPI, 
persist up to fairly high doping \cite{CarvaPime05a,CarvaPime05b}.   
Hence, the most important part of the interpaly between 
short-range AFM
and short-range EPI survives at moderate dopings.  

The hole also interacts with
dispersionless optical phonons $\Omega$ via
a short range Holstein coupling $\gamma$ \cite{Ishihara1,Ishihara2}
$$
\hat{H}^{\mbox{\scriptsize EPI}} = \Omega \sum_{\bf k} b_{\bf
k}^{\dagger} b_{\bf k} 
$$
\begin{equation} 
+ N^{-1/2} \gamma \sum_{\bf k , q} \left[
h_{\bf k}^{\dagger} h_{\bf k-q} b_{\bf q} + h.c. \right] \;.
\label{e-ph}
\end{equation}
(Here $b_{\bf k}$ and $h_{\bf k}$ are phonon and hole annihilation
operators, respectively.) We define the dimensionless coupling
constant $\lambda=\gamma^{2}/4t\Omega$ of the EPI. Our calculations
are done for parameters which are required to
reproduce the observed dispersions \cite{ZX95,Xiang_96} in
Sr$_2$CuO$_2$Cl$_2$ (SCOC) [$J=0.4t$, $t^{\prime}=-0.34t$,
$t^{\prime\prime}=0.23t$] and $\Omega=0.2t$.
The energies of the kink are 50-80 meV \cite{Lanzara01} 
and those of active phonons are 40-70 meV. Hence, with typical hopping  
amplitude $t \approx 0.4 eV$ one finds $0.12 \le \Omega/t \le 0.2$.
In further calculations we use spin-wave representation  
\cite{LiuMa92} and take into account only phonon-phonon vertex 
corrections. 
It was proved that the above approximations are reliable 
for the above parameters \cite{LiuMa92,GuRo06}.

The polaron picture for the finite doped case explains the
dichotomy between nodal and antinodal points \cite{ZXSch97,Zhou04}.
The fact that the lifetime of the lowest peak in ARPES
is large at the nodal point and small for the antinodal one, 
is a consequence of the contrasting nature of quasi-particle
broadening by EPI at different energies and momenta in the 
intermediate and weak coupling regimes. 
If the energy of the hole  at momentum ${\bf k}$, 
measured from the ground state of the Mott insulator in the nodal point
is smaller than the phonon frequency
$\Omega$, the decay of the quasiparticle by phonon emission is
forbidden by energy conservation law leading to the sharp peak.
On the other hand, when the energy is larger than the phonon
frequency (antinodal points, as an example) the real decay processes
by phonon emission are allowed, causing significant line broadening
even at moderate EPI coupling constants. For example, as it is seen
from the calculated momentum dependence of the Lehman function
(Fig.~\ref{fig:fig1}a,b), the linewidth abruptly increases when the
quasi-particle dispersion crosses the phonon energy at
$[k-\pi/2]/\pi \approx 0.17$. We emphasize that the interaction of
holes with phonons is essential for the explanation of
experimentally found ``dichotomy''. Although a broadening at general
momentum ${\bf k}$ is present even in pure $tt't''-J$ model, it is
considerably smaller than the experimental one and, more
important, does not show a sharp threshold-like increase above $\sim
50-70$meV. 
Of course one can not rule out inhomogeneous source of broadening of 
antinodal quasiparticles (see, e.g. \cite{Alldredge}) though the 
phonon mechanism is one of possible candidates to explain the ``dichotomy''. 
\begin{figure*}
\begin{center}
\includegraphics[width=19cm]{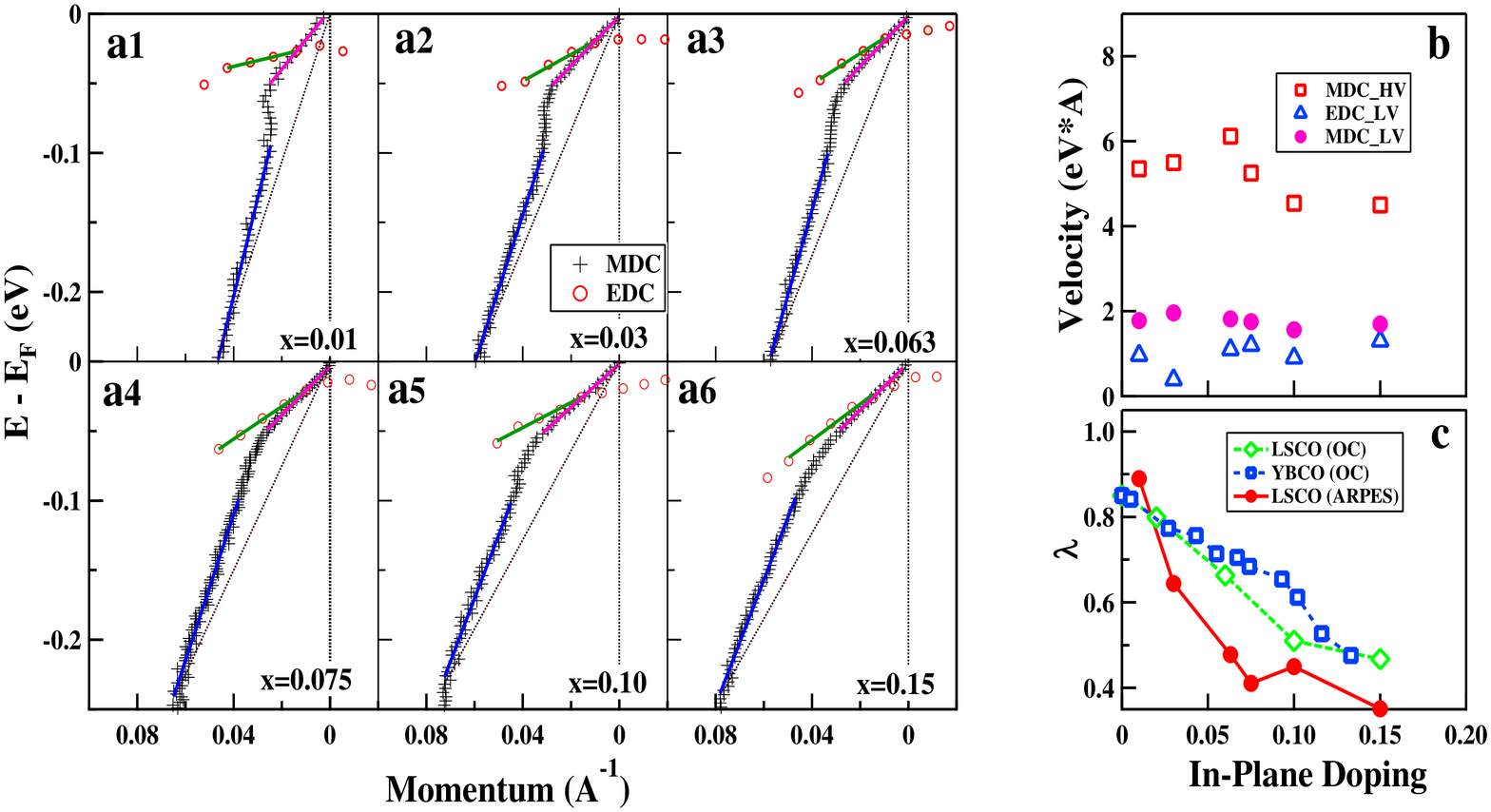}
\end{center}
\caption{\large (a). Energy-momentum dispersions for LSCO
with different dopings, using both EDC and MDC methods.  
The MDC low (high) energy velocity $V_{low}$ ($V_{high}$) is obtained 
by fitting MDC dispersion at  binding energy 0$\sim$50meV  
(100$\sim$250meV) using a linear line.   
(b). Low and high-energy velocities as a function of
doping obtained from MDC and EDC dispersions. (c). The EPI
coupling constant obtained from EDC low-energy velocity and high
energy velocity by empirical scaling relation (\ref{lalala}) 
$\lambda =
\sqrt{(V_{high}^{EDC}(\lambda)-V_{low}^{EDC}(\lambda))
/V_{low}^{EDC}(\lambda})/\sqrt{20}$
(filled circles). 
Empty squares and diamonds are coupling constants 
$\lambda^{\mbox{\scriptsize (OC)}}$ obtained from the analysis 
of the optical conductivity \cite{tJOpt08} on weakly doped LSCO and 
YBa$_2$Cu$_3$O$_{6+x}$ (YBCO), respectively. 
Only the ratio of $\lambda^{\mbox{\scriptsize (OC)}}(x)$ at finite doping 
$x$ and of that $\lambda^{\mbox{\scriptsize (OC)}}(x=0)$ at zero doping is 
determined in Ref.~\cite{tJOpt08} and we take 
$\lambda^{\mbox{\scriptsize (OC)}}(x=0)=0.85$ to match results obtained from 
optical and ARPES measurements as close as possible.}
\label{fig:fig2}
\end{figure*}

The low-energy dispersion (close to Fermi level $\varepsilon_{F}$), where
quasiparticles are sharp, is separated from the high-energy part by
a kink, where the velocity of the quasiparticle abruptly changes
\cite{Zhou03,Bogdan00,Lanzara01}. Experiments show that the
high-energy velocity above the bosonic mode $\Omega$ exhibits a
strong doping dependence, while the doping dependence of the
low-energy velocity is weaker \cite{Zhou03}.
Fig.~~\ref{fig:fig2}a shows the experimentally extracted
dispersions from LSCO \cite{Zhou03}.  
These data were obtained using the  
momentum distribution curve (MDC) method, giving the so-called universal 
low-energy velocity behavior where the velocity below the kink energy is
found to be largely insensitive to doping. 
In the framework of the MDC analysis one traces 
the momentum dependence of intensity at a fixed energy. 
Then, the point of maximal intensity is considered to be an energy-momentum
point representing the dispersion of the quasiparticle.    

DMC method  is capable of obtaining 
the energy distribution curves (EDC) when one traces the energy dependence of
the intensity keeping the momentum fixed.
The reason is that the theoretical spectra here are obtained as analytic 
continuation to the real frequencies of imaginary time Green functions 
calculated by the DMC method at fixed momenta. 
Green functions calculated at different momenta possess slightly different
distributions of statistical errorbars over the imaginary time.    
This difference is considerably amplified by the very high sensitivity of 
analytic continuation method to the distribution of statistical errors.
Hence, in contrast to EDC method the MDC method is not reliable in the 
framework of the theoretical methodology used here.

In order to compare directly experimental results with the calculated EDC
dispersion, we also re-analyze the low energy
data using the traditional EDC method. The extracted MDC and EDC
dispersions are plotted in Fig.~\ref{fig:fig2}a and the
corresponding low- and high-energy velocities are summarized in
Fig.~\ref{fig:fig2}b. 
The MDC and EDC dispersion curves are very similar in the weak
coupling limit, except for energies which are very close to phonon 
frequency \cite{Gun08}.
However, MDC and EDC curves in the strongly interacting systems, 
as it was already observed before for bi-layer manganite \cite{NormanMDC}, 
are very different.
The difference is especially large at low energies while the high energy 
data are rather insensitive to the type of analysis.

According to the analysis of the optical absorption \cite{tJOpt08},
the EPI is enhanced with underdoping. 
Then, as is expected  in the Migdal-Eliashberg picture,
the velocity at low energy should get smaller with underdoping while 
the high energy 
``bare'' velocity above the phonon frequency should not change. 
The observed behavior is different.
In the framework of the experimental MDC analysis (Fig.~\ref{fig:fig2}b), 
similar to the theoretical EDC results (Fig.~\ref{fig:fig1}c), 
the high-energy velocity {\it increases} with decreasing doping 
(Fig.~\ref{fig:fig2}b).
While understanding the differences between EDC and MDC is a matter 
to be further explored, we compared EDC experimental result with 
EDC theoretical data which are only ones available from calculations done 
by DMC method. 

In contrast to the Migdal-Eliashberg picture, 
a polaron picture in the intermediate coupling
regime can give a consistent explanation of these unique features.
We calculated ARPES spectra for different values of $\lambda$ along
the $(0,0)$ - $(\pi , \pi)$ nodal direction (Fig.~\ref{fig:fig1}c).
Indeed, the energy of the ground state $E(\pi/2,\pi/2)$ of the Hamiltonians 
(\ref{tJ}-\ref{e-ph}) depends on the coupling constant $\lambda$ 
(Fig.~\ref{fig:fig1}a,b) and, thus, the energies of the kinks at 
different $\lambda$s do not match to each other in the theoretical 
calculations.
The energy onset in experiment is set in a different way.
In the experimental data, presented in Ref.~\cite{Zhou03}, the binding energy 
is counted from the Fermi energy $\varepsilon_{F}$
and the origin of momentum is set to the Fermi momentum $k_{F}$
where the quasi-particle dispersion crosses the Fermi energy.
In this manner the energy of the kink in experimental data 
is roughly the same for all dopings \cite{Zhou03}.  
Hence, to compare the results of our theoretical calculations
with the experimental data for each $\lambda$ in Fig.~\ref{fig:fig1}(c)
we applied a constant energy shift to the theoretical results in order to 
match the energies of the quasi-particle for momenta just at the kink 
position.
It is clear that the theoretical high-energy $V_{high}^{EDC}(\lambda)$ and
low-energy $V_{low}^{EDC}(\lambda)$ velocities
estimated just above and below the kink position, respectively,
 do not depend on the theoretical energy onset.  
However, both the high- and low-energy EDC velocities are dependent on coupling 
constant $\lambda$, representing the intrinsic quasiparticle
behavior. 

The calculated high-energy velocity (Fig.~\ref{fig:fig1}c) in terms of the 
polaron picture gives a qualitative explanation of the anomalous doping 
dependence of the high-energy velocity observed in LSCO provided one
assumes that the effective coupling constant $\lambda$ increases with
underdoping. 
Indeed, one can see that, in contrast to the Migdal-Eliashberg picture, 
the theoretical high energy velocity $V_{high}^{EDC}(\lambda)$ 
(Fig.~\ref{fig:fig1}c) just above the kink position increases with 
$\lambda$. 
Hence, according to experiment, the theoretical high energy velocity 
$V_{high}^{EDC}(\lambda)$ increases with underdoping.

Furthermore, from our theoretical results of the spectral function of
a single hole in the $tt't''-J-ph$ model (\ref{tJ}-\ref{e-ph}),
we can fit the dependence of the ratio 
$(V_{high}^{EDC}(\lambda)-V_{low}^{EDC}(\lambda)) 
/V_{low}^{EDC}(\lambda)$ on the effective coupling constant $\lambda$
by a quadratic empirical scaling relation:    
\begin{equation} 
\frac{(V_{high}^{EDC}(\lambda)-V_{low}^{EDC}(\lambda))}
{V_{low}^{EDC}(\lambda)} \approx 20 \lambda^2 \; .
\label{la_fit}
\label{lalala}
\end{equation}
The doping dependence of the effective EPI constant $\lambda$ 
of LSCO estimated from the comparison of the above theoretical 
relation with experimental data is shown in Fig.~\ref{fig:fig2}c. 
Note, that the values of $\lambda$ obtained from ARPES are consistent
with data derived from the doping dependence of the optical absorption
\cite{tJOpt08} and other methods which take into account the strong 
interplay between EEI and EPI \cite{Gun08,UFN09}. 
We checked that the basic characteristics of the kink are mostly 
defined not by the range of EPI but by the resonance between 
the energy of the phonon and quasiparticle.  
Therefore, the data in  Fig.~\ref{fig:fig2}c are also valid for the 
Fr\"{o}hlich type of coupling.
We note that our parameters are always in the large-polaron 
regime because EPI is less than the critical coupling 
$\lambda_c=0.6$ for given model \cite{MN2006}.
The restriction of our analysis to a single phonon mode does not 
influence the relation (\ref{la_fit}) because fine features, 
caused by interaction with multiple modes, do not change 
the gross shape of the kink \cite{Zhou2005}.

Note here that the obtained effective $\lambda$s are reasonably of 
the order or less than unity even in the strongly underdoped regime where it is maximal.
This is in sharp contrast to the naive application
of the Migdal-Eliashberg relation $(V_{high}-V_{low})/V_{low}
=\lambda$ which would give an unphysically large number on the order 
of $\sim 10$ at low dopings.
It is clear that the Migdal-Eliashberg approach underestimates the effects
of EPI because the vertex corrections are neglected which, in turn,
leads to an overestimated value of the effective $\lambda$.
Indeed, even generalized Migdal-Eliashberg approach, including 
electron-electron correlation effects \cite{Mazur,Johnston}, predicts 
that $\lambda \approx 1.2$ in the optimal doping, which is 3 times larger
than our result obtained with full vertex corrections included. 

The values of $\lambda$ at low dopings, obtained
in this paper from the kink angle, coincide with that obtained
from the linewidth and distance of Franck-Condon peak from the 
chemical potential in undoped La$_2$CuO$_4$ compound \cite{Rosch}.
According to Fig.~1f in \cite{Lanzara01}, the kink angle is a rather universal
function and, hence, the performed analysis applies also to 
La$_{2-y-x}$Nd$_y$Sr$_x$CO$_4$, Bi$_2$Sr$_2$CuO$_6$ and 
Bi$_2$Sr$_2$CaCu$_2$O$_8$.  

In the vicinity of the kink (Fig.~\ref{fig:fig1}a) high resolution 
Lehman spectral functions consist of the ground
state peak and several phonon sidebands. The origin of the kink
is the abrupt transfer of spectral weight from the low energy peak to
the phonon sidebands when the momentum is increased (Fig.~\ref{fig:fig1}a).
The rate of transfer with momentum increase strongly
depends on the EPI coupling constant and, thus, the velocity just
above the kink energy is very sensitive to the EPI. 

Note, that the high energy velocity $V_{high}$ is more 
sensitive to EPI change than the renormalized low
energy velocity $V_{low}$ (Fig.~\ref{fig:fig1}c). We note that the
domain of the fast high energy velocity $V_{\mbox{\scriptsize
high}}$ is restricted to an energy range of a few phonon frequencies
$\Omega$ and to a small fraction of the Brillouin zone. Then,
according to our calculations, at higher momenta/energies dispersion
of the hump returns to the position of quasi-particle band
noninteracting with phonons. 
The return of the quasi-particle
dispersion to the unrenormalized one is typical also for the weak
coupling theory and for the Migdal-Eliashberg approach neglecting the 
vertex corrections. 
However, in the intermediate coupling regime the energy/momentum 
domain of large velocity in exact DMC method is considerably wider 
than that obtained in a weak coupling approach.

Several phonon sidebands survive above the momentum of
kink (Fig.~\ref{fig:fig1}a) though they are not seen (Fig.~\ref{fig:fig1}b)
if the spectrum is broadened more than it follows from the 
Hamiltonian  (\ref{tJ}-\ref{e-ph}).
Indeed there are numerous sources of decay, such as coming from 
hole-hole interaction, impurities, surface roughness etc., which
do not enter the Hamiltonian (\ref{tJ}-\ref{e-ph}). 
The intensity map where numerical data are broadened
by Gaussian with the width $\sigma=0.1$ does not show multiple
sidebands but a single peak (Fig.~\ref{fig:fig1}b), 
which roughly corresponds to the energy of the highest intensity sideband.  
Comparing Figs.~\ref{fig:fig1} and \ref{fig:fig2},
the global features of the spectrum are well reproduced by our
polaron picture, which provides an alternative and complementary
approximation replacing Migdal-Eliashberg approach in the underdoped regime. 
Strictly speaking, general features of the evolution of the Lehmann 
spectral function across the phonon energy are not restricted to the 
$tt't''-J-ph$ model since similar phenomena should be observed for any 
polaronic model with intermediate strength of EPI
\cite{Hohen03}. 
What is novel here is that a similar scenario is realized in a more
complicated situation with a combination of Mott physics and EPI. 

\bigskip

{\bf In conclusion}, we propose a novel polaronic metal picture for
underdoped cuprates and analyze the ARPES on underdoped cuprates in
the framework of the polaronic scenario
based on t-J model plus EPI.
It is particularly important that the electron-lattice polaron is 
formed in the background of the antiferromagnetic correlations and the 
essential influence of the electron-phonon interaction on the spectra
is observed exclusively due to constructive interplay between 
electron-lattice and electron-electron interaction.
Indeed, strong effects of coupling of holes to the lattice vibrations
are observed only because the latter are highlighted by concomitant
interaction of holes with short-range antiferromagnetic correlations.
The latter ones persist up to moderate dopings even when the 
long-range antiferromagnetic order is destroyed.  
The doped holes form the small polaron in the undoped and heavily 
underdoped region.
Analysis of ARPES in this scenario shows that, 
in accordance with the results previously obtained from the analysis of optical 
absorption, effective EPI decreases gradually and reaches the
self-trapping crossover at optimal doping. 
The extended (large polaron) and localized (small polaron) states 
coexist around this crossover region, which appear in a momentum 
dependent way. 
This picture explains many experimental puzzles, through 
the realistic estimate of the coupling constant, such as the large
momentum-dependent broadening of a single hole spectral function,
the dichotomy of nodal and anti-nodal direction, and the
unconventional doping dependence of ARPES of underdoped cuprates.

\bigskip

We thank G.\ Sawatzky,
O Gunnarsson, O.\ R\"{o}sch, G.\ Khaliullin, D.\ Tessa's, A.\ V.\
Chubukov, V. Cataudella, G. De Filippis, and C.\ Varma 
for valuable discussions. 
ASM is supported by RFBR grant 10-02-00047a.
NN is supported by MEXT Grand-in-Aid No.20740167, 19048008, 
19048015, and 21244053, 
Strategic International Cooperative Program (Joint Research Type) from 
Japan Science and Technology Agency, and by the Japan Society for the 
Promotion of Science (JSPS) through its ``Funding Program for
World-Leading Innovative R \& D on Science and Technology (FIRST Program)''. 
KMS acknowledges the National Science Foundation (DMR-0847385) 
and the Air Force Office of Scientific Research 
(Award No. FA9550-11-1-0033). 
TPD and ZXS are supported by the US Department of Energy, Office of 
Basic Energy Sciences under contract No. DE-AC02-76SF00515.

\end{document}